\documentstyle[prd,aps]{revtex}
\input epsf.sty

\def\la{\mathrel{\mathpalette\fun <}}
\def\ga{\mathrel{\mathpalette\fun >}}
\def\fun#1#2{\lower3.6pt\vbox{\baselineskip0pt\lineskip.9pt
\ialign{$\mathsurround=0pt#1\hfil##\hfil$\crcr#2\crcr\sim\crcr}}}

\newcommand{\be}{\begin{eqnarray}}
\newcommand{\ee}{\end{eqnarray}}

\preprint{\vbox{\hbox{NORDITA 98/39-HE}}}

\title{Electron Self-Energy in Strong Magnetic Field: Summation\\
of Double Logarithmic Terms}

\author{V. P. Gusynin\cite{email1}$^{1}$ and
A. V. Smilga\cite{email2}$^{2}$}

\address{$^{1}$Bogolyubov Institute for Theoretical Physics,
252143 Kiev, Ukraine\protect\\
$^{2}$NORDITA, Blegdamsvej 17, Copenhagen $/\!\!\!\!O$, DK-2100, Denmark
\footnote{Permanent address:
ITEP, B. Cheremushkinskaya 25, Moscow 117259, Russia.}
}

\date{July 22, 1998}
\draft
\newcommand{\grpicture}[1]
{
    \begin{center}
        \epsfxsize=300pt
        \epsfysize=200pt
        \vspace{5mm}
        \parbox{\epsfxsize}{\epsffile{#1.eps}}
        \vspace{5mm}
    \end{center}
}

\begin{document}

\maketitle

\begin{abstract}
We study the electron self--energy in a strong magnetic field when
the parameter $\eta\equiv (\alpha/2\pi) \ln^2 (eB/m^2_0) \sim 1$
and explore the transition between the perturbative regime
$\eta<<1$ and the nonperturbative massless QED regime $\eta>>1$
where chiral symmetry is broken spontaneously and electrons acquire
the dynamically induced mass.
\end{abstract}


\section{Introduction.}

Recently, it has been shown that, in the massless QED, the chiral
symmetry of the lagrangian is broken spontaneously when the system is placed
in an external magnetic field \cite{GMSprd,GMSnph,Leung,Hong}. Spontaneous
chiral symmetry breaking
is associated with the appearance of the gap in the electron spectrum
\cite{GMSnph}
\begin{equation}
\label{gap}
m_{\rm dyn}(B) \propto \sqrt{eB} \exp\left\{-{\pi\over
2}\sqrt{\frac{\pi}{2\alpha}}
\right\}\ ,
\end{equation}
the formation of the fermion condensate $<\bar e e>_B \sim eB
m_{\rm dyn}(B)$ \cite{SS}, and the appearance of the massless ``pion''.

\begin{figure}
\begin{center}
        \epsfxsize=300pt
        \epsfysize=60pt
        \vspace{5mm}
        \parbox{\epsfxsize}{\epsffile{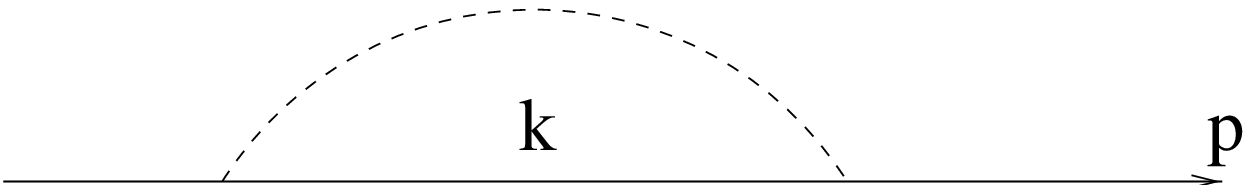}}
        \vspace{5mm}
    \end{center}
\caption{}
\end{figure}

On the other hand, in the real World with a finite physical electron mass
$m_0$ and when an external magnetic field is weak enough, we are in
the perturbative regime and the shift of the pole of the electron propagator
can be found by calculating the one--loop graph in Fig. 1
 in the presence of the field. Assuming $eB/m_0^2 \gg 1$, but $(\alpha/2\pi)
 \log^2(eB/m_0^2) \ll 1$, the result is \cite{jancov}
 \be
 m(B) \ =\ m_0 \left[ 1 + \frac \alpha {4\pi} \log^2 \frac {eB}{m_0^2}
 + \cdots \right]
\label{1loop}
  \ee
 In this paper, we explore the region when the dimensionless parameter
$\eta = (\alpha/2\pi)  \log^2(eB/m_0^2)$ is of order 1. We perform
an accurate summation of all leading double logarithmic
 contributions $\propto \left[\alpha \log^2 (eB/m_0^2) \right]^n$ and show that
 the result of the summation is a nontrivial function which has a {\rm singularity}
 at some value of   $\eta$. {\rm This} signalizes the breakdown of
the perturbation theory and the transition to the nonperturbative
regime involving spontaneous breaking of chiral symmetry. We solve
then a self--consistent Hartree--Fock equation and find the
effective mass dependence in the whole region of the parameter
$\eta$ which
 interpolates smoothly between perturbative result (\ref{1loop}) (and its
 leading logarithmic version) at small $\eta$ and the
 nonperturbative asymptotics (\ref{gap}) at  large $\eta$.

\section{Gross -- Neveu model.}

Before going over to four dimensions, let us  carry out the same
program in the two--dimensional Gross--Neveu model
\cite{GrossNeveu}. The model is  exactly solvable for large number
of fermions and its dynamics is very well understood.  We will see
later that $QED_4$ in an external magnetic field and the
Gross--Neveu model have very much in common and the physics of the
transitional region in these two theories is essentially the same.

The lagrangian density of the model reads
 \be
\label{GN}
{\cal L} \ = \ \sum_i \bar \psi_i (i \not{\!\partial} - m_0)
\psi_i +\frac{g^2}{2} \left[ (\sum_i \bar \psi_i\psi_i)^2-(\sum_i \bar\psi_i
\gamma_5\psi_i)^2\right],
\label{GNlagrangian}
\ee
where the fermion field carries an additional "color" index
$i=1,\dots N$. In the absence of bare mass $m_0$ the lagrangian
(\ref{GNlagrangian}) possesses $U(1)$ chiral symmetry $\psi_i\to
e^{i\theta\gamma_5}\psi_i$.  In two dimensions, the coupling $g$ is
dimensionless. We will assume it to be small.  When the number of
colours $N$ is large, the perturbative series
 involves only the so called daisy or, better to say, ``cactus''
graphs (see Fig. 2); all other graphs are
suppressed as inverse powers of $N$.

\begin{figure}[t]
\grpicture{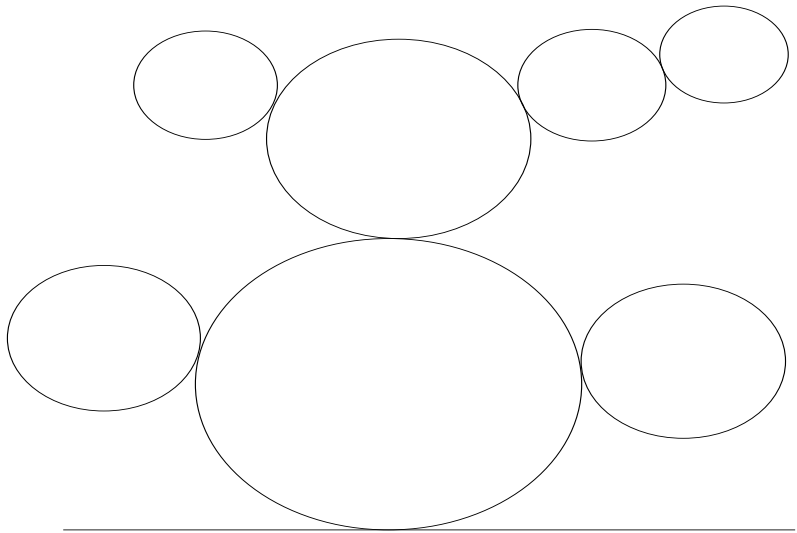}
\caption{A cactus graph.}
\end{figure}

The 1--loop correction to the bare fermion mass $m_0$ is given by the
graph in Fig. 3a. A trivial calculation gives
  \begin{equation}
m \ =\ m_0 \left[ 1 + \alpha \log\frac{\Lambda}{m_0} + o(\alpha)
\right],
\label{GN1loop}
  \end{equation}
where $\alpha=g^2N/\pi$ and $\Lambda$ is an ultraviolet cutoff.
Next, we can sum up all the leading logarithmic terms $\sim~[\alpha
\log (\Lambda/m_0)]^n$. They are given by the graphs in Fig. 3b.
The result is
 \begin{equation}
m^{LL}\  = \ \frac{m_0}{1-\alpha\log\frac{\Lambda}{m_0}}.
\label{1_ordersol}
\end{equation}

The leading logarithmic expression blows up at a finite cutoff which
signalizes the breakdown of the perturbation theory even in its
improved form.
 In this simple case, however, we can sum up not only the leading
logarithmic cactus graphs, but just {\it all} of them if writing down
the self--consistent gap equation
  \begin{equation}
m \ =\ m_0  + \alpha m\log\frac{\Lambda}{m},
\label{GNgapeq}
\end{equation}
  The solution of this equation is a smooth function of $\Lambda$.
When $\alpha \log(\Lambda/m_0)$ is small, it coincides with the
perturbative unimproved, and further, improved results
(\ref{GN1loop}), (\ref{1_ordersol}). When $\alpha \log(\Lambda/m_0)
\gg 1$, we find ourselves in a nonperturbative regime. In this case,
one can neglect the bare mass altogether, and the solution of
Eq.(\ref{GNgapeq})
is just

\begin{figure}
\grpicture{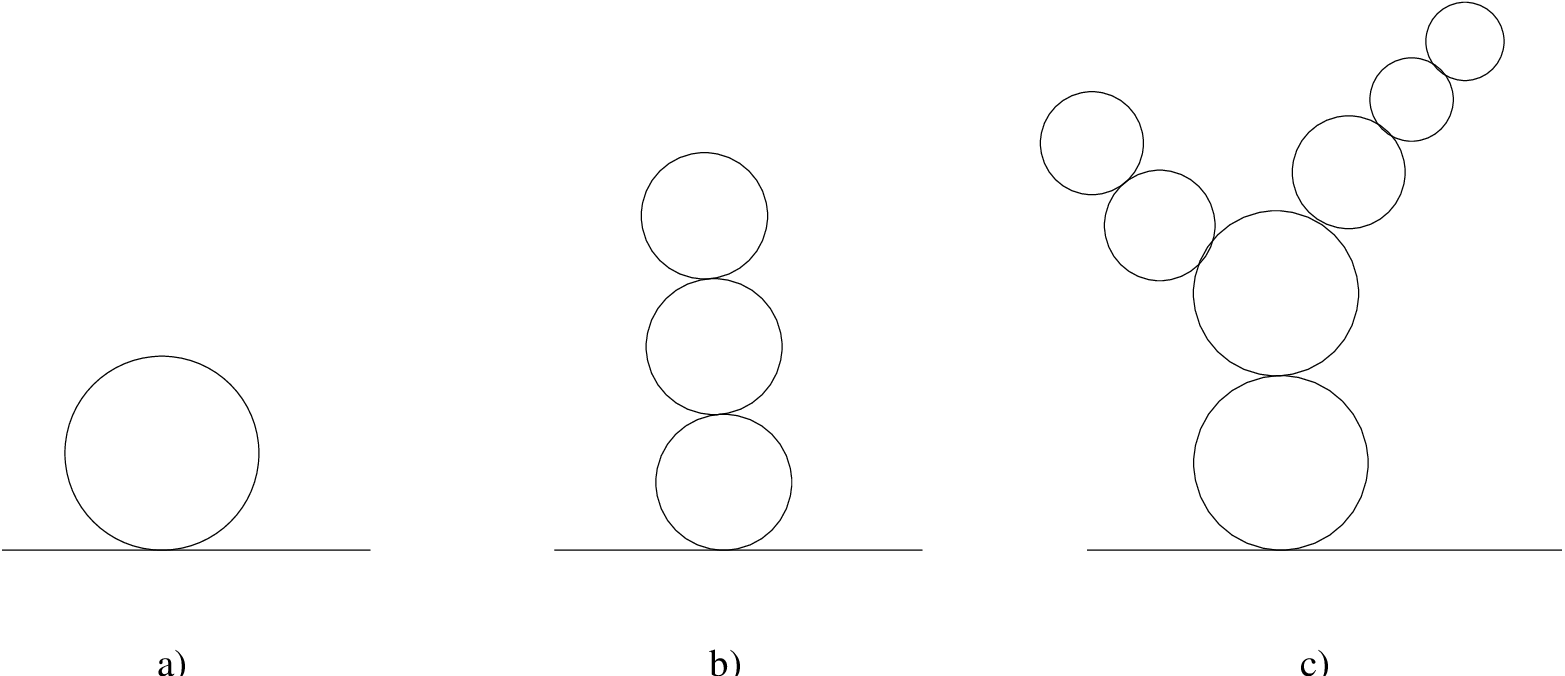}
\caption{Graphs describing mass renormalization in the Gross--Neveu
model in {\it a)} one loop, {\it b)} leading
logarithmic approximation, {\it c)} next-to-leading logarithmic
approximation.}
\end{figure}

  \begin{equation}
\label{GNgap}
m \ =\ \Lambda e^{-{1\over\alpha}}.
\end{equation}
 A finite gap (\ref{GNgap}) signalizes the spontaneous breakdown of
 the chiral symmetry in the massless Gross--Neveu model in the leading
 order in $1/N$ expansion \cite{GrossNeveu}\footnote{
The Mermin-Wagner-Coleman theorem \cite{MWCtheorem} forbids the
spontaneous breakdown of continuous symmetries in two dimensions.
The fermion dynamical mass $m$ is an order parameter of chiral
symmetry breaking only in leading order in $1/N$. In the
next-to-leading order approximation, spontaneous chiral symmetry
breaking is washed out by interactions of would--be Nambu-Goldstone
(NG) bosons. In fact, in this model we have the realization of the
Berezinsky--Kosterlitz--Thouless (BKT) phase: while chiral symmetry
is not broken in this phase, the fermions acquire dynamical mass
$m$, and the would--be NG boson transforms into a BKT gapless
excitation \cite{Witten}.
 }.

It is instructive to rewrite the transcendental equation
(\ref{GNgapeq}) in a differential form
\begin{equation}
\frac{dy}{d\xi}=\frac{\alpha y^2}{1+\alpha y},\qquad y(0)=1,
\label{GNdiffeq}
\end{equation}
where we introduced   $y=m/m_0$ and $\xi=\log{\Lambda/m_0}$. This
is nothing else as the standard renormalization group equation for
a mass. If we now expand the right hand side of (\ref{GNdiffeq})
keeping only the leading in $\alpha$ term, we come at the singular
solution (\ref{1_ordersol}).

In the next order in $\alpha$ we have the
equation which takes care of  all the leading and
next--to--leading logarithmic terms:
\begin{equation}
\label{a2y3}
y^\prime=\alpha y^2-\alpha^2y^3.
\end{equation}
In the diagram language, it corresponds to the summation of a
particular infinite set of cactus graphs with one chain (like in
Fig. 3b) or with two or more chains branched out of one and the same link
where a logarithm is lost (see Fig. 3c). The solution  tends
to a "fixed point" $y_* = 1/\alpha$ when $\xi\to\infty$.

In the order $\sim \alpha^3$, the equation summing leading,
next--to--leading, and next--to--next--to--leading logarithms reads
\begin{equation}
\label{a3y4}
y^\prime=\alpha y^2-\alpha^2y^3+\alpha^3y^4,
\end{equation}
In the graph language that corresponds to taking into account also the graphs
with 2 branchings.
The solution of Eq. (\ref{a3y4})  blows up at a finite $\xi$ like
it did in the leading order. In the still
next order $\sim
\alpha^4$, we obtain again the equation with (the same !) fixed point etc.

The solution to the exact equation (\ref{GNdiffeq}) does not have any fixed
point,
neither does it blow up at a finite $\xi$.
We plotted it (for $\alpha = .1$)  together with the perturbative
approximations [Eq.(\ref{1_ordersol}]
and the solution to the equation (\ref{a2y3})] and the nonperturbative
asymptotics (\ref{GNgap}) [
$y_{\rm as}(\xi) = \exp(\xi-1/\alpha)$ in our variables] in Fig. 4.

\begin{figure}
\grpicture{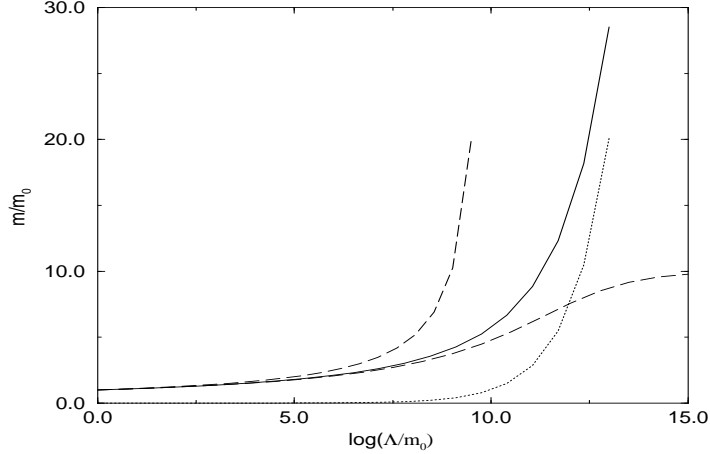}
\caption{Fermion mass in the Gross--Neveu model (solid line). The
upper and lower dashed lines are the result of summation of, correspondingly,
the leading and the
next--to--leading perturbative logarithms and the dotted line is the
nonperturbative asymptotics.}
\end{figure}

\section{QED in a magnetic field}

To begin with, let us spend some time to explain how the leading
logarithmic 1--loop result (\ref{1loop}) can be derived. The experience and insights
acquired will be useful when handling
higher--loop contributions.

To calculate the graph in Fig. 1, we need to know the electron
Green function in a magnetic field. The explicit expression has
been found by Schwinger \cite{Schwinger}. First of all, we can
write
 \be
\label{phase}
G_B(x,y) \ =\ \exp\left\{ ie \int_y^x A_\mu(\xi) d\xi_\mu \right\}
\int \frac {d^4k}{(2\pi)^4}e^{-ik(x-y)} G_B(k),
 \ee
where the integral in the phase factor is done along the straight
line connecting $x$ and $y$. Generally, the momentum space fermion
propagator $G_B(k)$ is given by a complicated integral over the
proper time. It can be expanded in terms of Landau levels (the
magnetic field is directed along the third axis) \cite{Chod}
\be
G_B(k)=i\exp\left(-\frac{k^2_\bot}{eB}\right)\sum\limits_{n=0}^\infty
(-1)^n\frac{D_n(eB,k)}{k_\parallel^2-m_0^2-2eBn}
\label{LLdecomp}
\ee
with
\be
D_n(eB,k)=(\hat k_\parallel+m_0)\left[(1-i\gamma^1\gamma^2)
L_n\left(\frac{2k^2_\bot}{eB}\right)-(1+i\gamma^1\gamma^2)
L_{n-1}\left(\frac{2k^2_\bot}{eB}\right)\right] +4\hat k_\bot
L_{n-1}^1\left(\frac{2k^2_\bot}{eB}\right),
\ee
where $L_n^\alpha$ are the generalized Laguerre polynomials,
$k_\parallel$ and $k_\bot$ are longitudinal and transverse momentum
components, $\hat k_\parallel=k^0\gamma^0-k^3\gamma^3$, $\hat
k_\bot=k^1\gamma^1+k^2\gamma^2$, $k_\|^2 = k_0^2 - k_3^2$,
$k_\bot^2 = k_1^2 + k_2^2$.

 Fortunately, as we will see later, we do not need the full
expression (\ref{LLdecomp}) for our purposes, but only its
asymptotic form in the region of small momenta $k \ll \sqrt{eB}$.
In this region, it suffices to retain only the lowest Landau levels
(LLL) in the spectral decomposition of the Green function which
acquires then a very simple form
\be
\label{LLL}
G_B(k) \ =\ ie^{-k_\bot^2/eB} \ \frac {\hat k_\| + m_0}{k_\|^2 - m_0^2}
\ (1 - i\gamma^1 \gamma^2).
\ee
Basically, this Green's function describes free motion in the
longitudinal direction of the states with $\sigma = {\bf \Sigma
B}/|{\bf B}| = -1$  (the factor $1 - i\gamma^1 \gamma^2$ filters
out the states with wrong sign of the spin projection $\sigma$).
The LLL
approximation is justified when both $k_\bot^2$ and
$k_\|^2$ are much smaller than $eB$. Thereby, it is more consistent
to set the form factor $ e^{-k_\bot^2/eB}$ (appearing due to a
finite size of the Landau orbits) to 1, but we should keep in mind
that all the momenta integrals should be cut off at $k_\bot^2,
k_\|^2\la eB$.

Substituting Eqs.(\ref{phase}) and (\ref{LLL}) in the Feynman
integral, taking some care when handling the phase factors $\propto
\exp\left\{ ie \int A_\mu d\xi_\mu\right\}$, and performing the Wick
rotation, the 1--loop expression for the euclidean mass operator
acquires the form
\footnote{We have chosen the  Feynman gauge, but the mass defined as the
position of the pole of  electron propagator at zero momentum should not
depend on the gauge
and it does not \cite{GMSprd,GMSnph,Ferrer}.}

  \be
 \label{m1int}
M_1(p_\|^2) \ =\ \frac {\alpha m_0}{2\pi} \int^{eB} \frac {d^2
k_\|}{k_\|^2 + m_0^2} \int^{eB} \frac {dk_\bot^2}{k_\bot^2 + ({\bf
k}_\| - {\bf p}_\|)^2}.
 \ee
We assumed here that the external electron momentum $p$ is purely
longitudinal. One can show that the dependence of $M(p_\|^2,
p_\bot^2)$ on
$p_\bot^2$ is weak when $p_\bot^2 \ll eB$.

The integral (\ref{m1int}) is double logarithmic receiving a main
contribution from the region $m_0^2 \ll k_\|^2 \ll k_\bot^2 \ll
eB$. {\rm That} justifies the use of the approximate  LLL
expression (\ref{LLL}). For $k \ga \sqrt{eB}$, the LLL
approximation is not valid, but the corresponding contribution in
the mass operator would not involve a large logarithm
$\log(eB/m_0^2)$. To find the mass shift (\ref{1loop}), it suffices
to set $p_\|^2 = m_0^2$ in Eq.(\ref{m1int}) and calculate the
integral with double logarithmic accuracy. In the following, we
will need, however, also the off--shell mass operator with
arbitrary $p_\|^2$. An important remark  is that, if $p_\|^2 \gg
m_0^2$, the structure of the integrand in Eq. (\ref{m1int}) depends
on the relation between $k_\|$ and $p_\|$. There are two distinct
regions, both of them providing a double logarithmic contribution
in $M_1(p_\|)$.

{\it A)} $k_\|^2 \gg p_\|^2$. In that case, the integral is reduced
to
\be
\label{m1A}
M_1^A(p_\|^2) \approx \ \frac {\alpha m_0}{2\pi} \int_{p_\|^2}^{eB}
\frac {dk_\|^2}{k_\|^2}
\int_{k_\|^2}^{eB} \frac {dk_\bot^2}{k_\bot^2} \ =\
\frac{\alpha m_0}{4\pi} \log^2 \frac {eB}{p_\|^2}
\ee

{\it B)} $m_0^2 \ll k_\|^2 \ll p_\|^2$. We have
 \be
\label{m1B}
M_1^B(p_\|^2) \approx \ \frac {\alpha m_0}{2\pi} \int_{m_0^2}^{p_\|^2}
\frac {dk_\|^2}{k_\|^2}
\int_{p_\|^2}^{eB} \frac {dk_\bot^2}{k_\bot^2} \ =\
\frac{\alpha m_0}{2\pi} \log\frac {p_\|^2}{m_0^2} \log\frac {eB}{p_\|^2}
 \ee

The sum of two contributions (\ref{m1A}) and (\ref{m1B}) is
 \begin{equation}
M_1(p_\|^2)=\frac{\alpha m_0}{2\pi}\left[Ll - {1\over 2}L^2\right],
\qquad L=\log\frac{eB}{p_\|^2},\quad l=\log\frac{eB}{m_0^2}.
\end{equation}

\begin{figure}
\begin{center}
        \epsfxsize=300pt
        \epsfysize=100pt
        \vspace{5mm}
        \parbox{\epsfxsize}{\epsffile{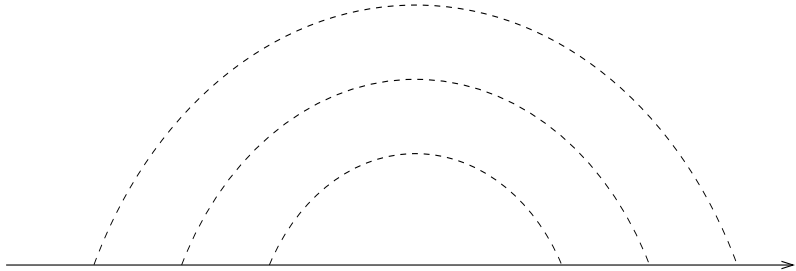}}
        \vspace{5mm}
    \end{center}
\caption{A rainbow graph.}
\end{figure}

Let us consider now higher loop contributions. The leading logarithmic
terms come from the rainbow graphs depicted in Fig. 5. The n-th order
contribution for the mass operator is obtained from the contribution
of the (n-1)-th order by the virtue of the recurrent relations
  \be
 \label{recur}
M_n(p_\|^2) \ =\ \frac {\alpha }{2\pi} \int^{eB} M_{n-1}(k_\|^2)
\frac {d^2 k_\|}{k_\|^2 + m_0^2} \int^{eB} \frac
{dk_\bot^2}{k_\bot^2 + ({\bf k}_\| - {\bf p}_\|)^2}.
 \ee
This is nothing else, of course, as the n-th approximant of the
integral equation
 \be
 \label{inteq}
M(p_\|^2) \ =\ m_0 + \frac {\alpha }{2\pi} \int^{eB} M(k_\|^2) \frac {d^2
k_\|}{k_\|^2 + m_0^2} \int^{eB} \frac {dk_\bot^2}{k_\bot^2 + ({\bf
k}_\| - {\bf p}_\|)^2}
 \ee
[ $M(p_\|^2) = m_0 + M_1(p_\|^2) + \cdots $].

Again, the leading logarithmic contributions in Eqs. (\ref{recur}),
(\ref{inteq})
come from two distinct kinematic regions:
$eB \gg k_\bot^2 \gg k_\|^2 \gg p_\|^2 \gg m_0^2$ and
$eB \gg k_\bot^2 \gg p_\|^2 \gg k_\|^2 \gg m_0^2$. Actually, as far as
the leading logarithmic terms are concerned, the equation
(\ref{inteq}) can be greatly simplified:
\begin{equation}\label{integraleq}
  M(p_\|^2)=m_0+\frac{\alpha}{2\pi}\left[\log\frac{eB}{p_\|^2}\int
  \limits_{m_0^2}^{p_\|^2}\frac {dk_\|^2}{k_\|^2} M(k_\|^2) +
\int\limits_{p_\|^2}^{eB}\frac {dk_\|^2}{k_\|}  \log\frac{eB}
  {k_\|^2} M(k_\|^2)\right].
\end{equation}
  This equation is easy to solve. One way to do it is to find the
  subsequent approximants
\begin{eqnarray}
 \label{Mexpan}
    M_2(p_\|^2)&=& \left(\frac{\alpha}{2\pi}\right)^2 m_0\left[{1\over 24}
L^4-{1\over 6}L^3l+{1\over 3}Ll^3\right],\nonumber\\
    M_3(p_\|^2)&=&\left(\frac{\alpha}{2\pi}\right)^3 m_0\left[-{1\over 720}
L^6+{1\over 120}L^5l-{1\over 18}L^3l^3+{2\over 15}Ll^5\right], \ \ldots
\end{eqnarray}
and to sum them over. The shift of the electron pole mass
[it is defined as the solution of the dispersive equation $m =
M(m^2)$, but to the leading logarithmic order, we may just
 set $m = M(m_0^2) = M(L=l)$]
is then given by the series
  \begin{equation}
 m \ =\ m_0 \left[1+\frac{\alpha}{4\pi}\log^2\frac{eB}{m_0^2}+{5\over
24}\left(\frac{\alpha}{2\pi}\log^2\frac{eB}{m_0^2}\right)^2+
{61\over
720}\left(\frac{\alpha}{2\pi}\log^2\frac{eB}{m_0^2}\right)^3 + \cdots
\right] \ .
\label{masscorr}
  \end{equation}

   A more nice and easy way to handle  Eq.(\ref{integraleq}) is to notice
that it is equivalent to the following second order differential equation:
      \be
\frac{d}{dp_\|^2}\left(p_\|^2\frac{dM(p_\|^2)}{dp_\|^2}\right)+\frac{\alpha}{2\pi}
\frac{M(p_\|^2)}{p_\|^2} \ =\ 0 \ ,
\label{diffeq}
  \ee
with the boundary conditions
\begin{equation}
\frac{dM(p_\|^2)}{dp_\|^2}\Bigr|_{p_\|^2=m_0^2}=0,\qquad
M(p_\|^2)\Bigr|_{p_\|^2 = eB} = m_0 \ .
\label{boundcond}
\end{equation}
The solution of Eqs.(\ref{diffeq}),(\ref{boundcond}) is
\begin{equation}
M(p_\|^2)=m_0\frac{\cos\left(\sqrt{\frac{\alpha}{2\pi}}\log
\frac{p_\|^2}{m_0^2}\right)}{\cos\left(\sqrt{\frac{\alpha}{2\pi}}
\log\frac{eB}{m_0^2}\right)}.
\label{massfunction}
\end{equation}
The pole  mass of the electron [defined as $m=M(m_0^2)$] is
\begin{equation}
m=\frac{m_0}{\cos\left(\sqrt{\frac{\alpha}{2\pi}}\log\frac{eB}{m_0^2}
\right)}.
\label{fullpertmass}
\end{equation}
The expansion of these expressions in $\alpha$ reproduces, of
course, Eqs. (\ref{Mexpan}) and (\ref{masscorr}).
 Note that Eq.(\ref{fullpertmass}) is
different from an analogous formula obtained in \cite{skobelev}
which had the form
\begin{equation}
m=m_0\exp\left(\frac{\alpha}{4\pi}\log^2\frac{eB}{m_0^2}
\right) .
\label{skobelmass}
\end{equation}
 The terms $\sim \alpha$ in Eqs.(\ref{fullpertmass}) and
(\ref{skobelmass}) coincide with each other and with the Jancovici
result (\ref{1loop}), but the two--loop coefficients are already
different: we have obtained $5/24$ in the expansion
(\ref{masscorr}) instead of $1/8$ in \cite{skobelev}. Also the
behaviour of the functions (\ref{fullpertmass}) and
(\ref{skobelmass}) at large values of the argument is completely
different. When $m_0$ tends to zero, the  electron mass
Eq.(\ref{fullpertmass}), in contrast to Eq.(\ref{skobelmass}),
 runs into singularity at some finite $m_0$
signalizing the breakdown of perturbation theory [cf. Eq.
(\ref{1_ordersol}) !].

 The experience with the Gross--Neveu model taught us how to handle this
problem. We have to write a self-consistent gap equation which sums up
effectively all the relevant graphs. To this end, we just have to replace
 the mass $m_0$ in the integrand in Eq. (\ref{integraleq}) and,
correspondingly, in the argument of the cosine function in Eq.
(\ref {fullpertmass}) by the total mass $m(B)$.  We then come to
the equation
\footnote{Note that, in contrast to Eq. (\ref{GNgapeq}) in the Gross--Neveu
model which summed up {\it all} the perturbative graphs in the
large $N$ limit and was just exact, the status of Eq.
(\ref{dynmasseq}) is somewhat less certain. It is written in the
mean field (or Hartree--Fock) approximation neglecting the effects
due to  momentum dependence of the mass operator in the denominator
of the electron propagator. We assume, however, that the mean field
analysis is justified here as it is justified in many other
physical problems involving gap equations (like standard
superconductivity etc). Moreover, a numerical study of the
nonlinear gap equation \cite{GMSnph} confirms the formula
(\ref{gap}).}
\begin{equation}
m\cos\left(\sqrt{\frac{\alpha}{2\pi}}\log\frac{eB}{m^2}\right)=m_0.
\label{dynmasseq}
\end{equation}
The solution of this  equation is nonsingular in the limit $m_0 \to
0$ giving a nonzero dynamical mass (\ref{gap}) of  massless
electron in a magnetic field \cite{GMSnph}.

Like in the Gross--Neveu case, we can rewrite  Eq.(\ref{dynmasseq})
in a differential
renormalization-group-like form. Introducing the variable
$y=m/m_0$ and differentiating over $\xi = \log(eB/m_0^2)$ we obtain
\begin{equation}
y^\prime=\frac{\sqrt{\frac{\alpha}{2\pi}}y\sqrt{y^2-1}}{1+\sqrt
{\frac{2\alpha}{\pi}}\sqrt{y^2-1}} \ , \qquad y(0) = 1 \ .
\label{rgeqformass}
\end{equation}
We can consider the right hand side of Eq.(\ref{rgeqformass}) as an
analog of $\beta$-function for the "charge" $m/m_0$. The analysis is
exactly parallel to what we have done for the Gross--Neveu model. In the lowest
order in $\sqrt{\alpha}$, Eq.(\ref{rgeqformass}) is reduced to
\begin{equation}
y^\prime=\sqrt{\frac{\alpha}{2\pi}}y\sqrt{y^2-1}
\end{equation}
with the solution
\begin{equation}
y=\frac{1}{\cos\left(\sqrt{\frac{\alpha}{2\pi}}\log\frac{eB}{m_0^2}\right)},
\end{equation}
 which coincides with (\ref{fullpertmass}) and blows up at some finite
$m_0$ (or $B$). In the next order in $\sqrt{\alpha}$ we get
\begin{equation}
y^\prime = \sqrt{\frac{\alpha}{2\pi}} y\sqrt{y^2-1}\left[1-
\sqrt{\frac{2\alpha}{\pi}} \sqrt{y^2-1}\right].
\label{secondord}
\end{equation}
The right hand side of the last equation has a "fixed point" at
$y_0=\sqrt{1+\pi/2\alpha}$ to which the solution of
Eq.(\ref{secondord}) tends as $\log{eB/m_0^2}\to\infty$. If keeping
the term of order $(\sqrt\alpha)^3$ in the
expansion of right hand side of (\ref{rgeqformass}) the "fixed
point" disappears while reappearing  again in order
$(\sqrt\alpha)^4$. On the other hand, the exact equation
(\ref{rgeqformass}) does not have any fixed points (other than the
zero field point $y = 1$ ) and its right hand side behaves as $y/2$ at
large $y$ what means that the total electron mass behaves as
$m \propto \sqrt{eB}$ in a strong magnetic field.
The solution to the full Hartree--Fock equation (\ref{rgeqformass}),
its perturbative approximants, and the nonperturbative asymptotics
(\ref{gap}) are plotted in Fig. 6.
\begin{figure}
\grpicture{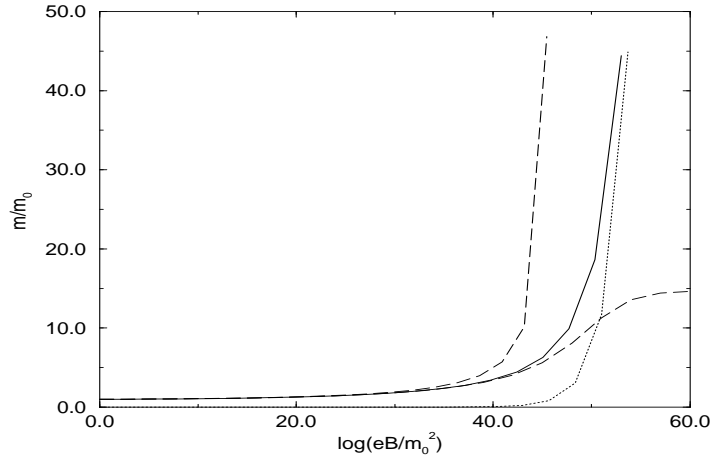}
\caption{Electron mass as a function of  magnetic field (solid line).
The dashed and dotted lines have the same meaning as in Fig. 4.}
\end{figure}

The main lesson to be learned from our analysis is that, for systems
 involving spontaneous
chiral symmetry breaking, the perturbative methods (even improved
by summation of the leading or subleading logarithms) have an
intrinsic barrier: one cannot go much further beyond the point
$\alpha \log (\Lambda/m_0) \sim 1$ for the Gross--Neveu model or
the point $(\alpha/2\pi) \log^2(eB/m_0^2) \sim 1 $ for $QED_4$ in
magnetic field: the perturbative approximants either blow up at
some small enough value of $m_0$ or display an unphysical fixed
point behaviour. In the region of very large $\Lambda/m_0$ for the
Gross--Neveu model and $eB/m_0^2$ for $QED$, the
effects due to explicit breaking of chiral symmetry in the
lagrangian become negligible and the correct results can be obtained
by solving a self--consistent mean field nonperturbative gap
equation.

In the real QED, the value of the parameter $\eta\simeq 1$ is reached
for fields of the order $\sim 10^{26}G$. We recall that strong
magnetic fields ($B\sim 10^{24}G$) might have been generated during
the electroweak phase transition \cite{Vachaspati}. It has been
speculated in \cite{GMSprd,GMSnph} that the character of
electroweak phase transition could be affected by generation of a
dynamical electron mass under such strong fields. On the other
hand, as is seen from Fig.6, the nonperturbative regime becomes
prevailing over the perturbative one for values of $\eta$ of the
order $2.35$ what corresponds to magnetic fields $\sim 10^{32}G$.
Ambj{\o}rn and Olesen \cite{Olesen} have claimed that even larger
fields, $\sim 10^{33}G$, would be necessary at early stages of the
Universe to explain the observed large-scale galactic magnetic
fields.

It was shown in \cite{Ferrer2} that the critical magnetic field,
required for the effect of chiral symmetry breaking to be
important, can be substantially decreased by taking into account
other interactions, for example Yukawa couplings. It would be
interesting to extend the results obtained in present paper to
Standard Model.\\

We thank V.M. Braun and V.A. Miransky for useful remarks. We would like to
acknowledge kind hospitality at the University of Bern where this
work has been started. The work has been completed when one of us
(A.V.S.) was staying at NORDITA and he is grateful to the members
of NORDITA for warm hospitality.
 The work of V.P.G. was partially supported by the grant
INTAS-93-2058-ext, by Swiss National Science Foundation grant
CEEC/NIS/96-98/7 IP 051219 and by Foundation of Fundamental
Researches of Ministry of Sciences of Ukraine under grant No.
2.5.1/003.
 The work of A.V.S. was done under the partial support of RFBR INTAS
grants 93-0283, 94-2851, RFFI grant 97-02-16131, CRDF grant
RP2-132, and Schweizerischer National Fonds grant 7SUPJ048716.

\end{document}